\documentclass[a4paper,11pt]{article}

\usepackage{jheppub} 

\usepackage[T1]{fontenc} 
\usepackage{amsmath,amssymb}
\usepackage{ifpdf}
\usepackage{ulem}

\newcommand{\Slash}[1]{{\ooalign{\hfil/\hfil\crcr\(#1\)}}}

\newsavebox{\boxa}

\newcommand{\eref}[1]{eq.~(\ref{#1})}

\title{\boldmath Flow Equation of N=1 Supersymmetric O(N) Nonlinear Sigma Model in Two Dimensions }

\author[a]{Sinya Aoki,}
\author[b,*]{Kengo Kikuchi,\note{Corresponding author.}}
\author[c]{Tetsuya Onogi}

\affiliation[a]{Center for Gravitational Physics, Yukawa Institute for Theoretical Physics, Kyoto University, Kyoto 606-8502, Japan}
\affiliation[b]{Maskawa Institute for Science and Culture, Kyoto Sangyo University, Kyoto 603-8555, Japan}
\affiliation[c]{Department of Physics, Osaka University, Toyonaka, Osaka 560-0043, Japan}

\emailAdd{saoki@yukawa.kyoto-u.ac.jp}
\emailAdd{kengo@yukawa.kyoto-u.ac.jp}
\emailAdd{onogi@phys.sci.osaka-u.ac.jp}

\abstract{We study the flow equation for the $\mathcal{N}=1$ supersymmetric $O(N)$ nonlinear sigma model in two dimensions,
which cannot be given by the gradient of the action, as evident from dimensional analysis.
Imposing the condition on the flow equation that it respects both the supersymmetry and the $O(N)$ symmetry, we show that the flow equation has a specific form, which however contains an undetermined  function of the supersymmetric derivatives $D$ and $\bar D$. 
Taking the most simple choice, we propose a flow equation for this model.  As an application of the flow equation, we give the solution of the equation at the leading order in the large $N$ expansion. The result  shows that the flow of the superfield in the model is dominated by  the scalar term, since the supersymmetry is unbroken in the original model.
It is also shown that the two point function of the superfield is finite at the leading order of the large $N$ expansion.
}

\begin{document} 
\hspace{7cm}YITP-17-28, MISC-2017-02, OU-HET-927\\
\maketitle
\flushbottom

\newpage 
\section{Introduction}
Recently, the method of the gradient flow~\cite{Luscher:2009eq, Luscher:2010iy, Luscher:2011bx} has been the focus of research in  lattice QCD.  The attractive properties of this method lies in the ultraviolet (UV) finiteness of the composite operators constructed from the fields at finite flow time $t$, which was shown in Ref.~\cite{ Luscher:2010iy, Luscher:2011bx}.  Based on this remarkable properties,  there have been a lot of applications to the studies of physical observables. In Ref.~\cite{Luscher:2013cpa}, nonperturbative renormalization and O(a) improvement of the axial current as well as the accurate determination of the chiral condensate were studied using the chiral Ward-Takahashi identities. In Ref.~\cite{Suzuki:2013gza}, the perturbatively renormalized energy momentum tensor in the continuum theory was extracted  from the composite operator at finite flow time $t$.
The energy momentum tensor so defined was  used to compute the equation of state in the Yang-Mills theory~\cite{Asakawa:2013laa}.  Nonperturbative renormalization of the energy momentum tensor was obtained 
in Ref.~\cite{DelDebbio:2013zaa}.

Being finite, the correlation fucntions at finite flow time are regularization independent and shares the same symmetry properties as the continuum theory.  
Therefore, even if the lattice regularization breaks the global symmetries such as chiral symmetry or translation/dilatation symmetry,  they 
can serve as natural probes of Ward-Takahashi identities.  

One of the most difficult but interesting problems is the lattice regularization of the supersymmetry and the nonperturbative studies of its dynamics. 
In the $\mathcal{N}=1$ supersymmetric pure Yang-Mills theory, naive lattice regularization breaks the supersymmetry, so that  
the fermion mass  parameter should be fine-tuned to recover the supersymmetry in the continuum limit~\cite{Curci:1986sm},  by imposing the chiral Ward-Takahashi identity. 
Even after this fine-tuning, it is still hard to confirm 
the recovery of the supersymmetry in the continuum limit via supersymmetric Ward-Takahashi identity 
in nonperturbative lattice simulations due to various systematic errors~\cite{Farchioni:2001wx}. 
This suggests that the fine-tuning program  for more general supersymmetric theories such as $\mathcal{N}=1$ supersymmetric QCD, where there exist five supersymmetry breaking parameters 
that are needed to be fine-tuned,  is hopelessly difficult.  
In view of this situation, it is natural to expect that once the flow equation is extended to supersymmetric theories,  they may help us to improve the study of supersymmetric lattice field theories. 

For historical reasons, the flow equation has been called the ``gradient'' flow equation, since the Yang-Mills flow equation, which is  a typical example of the flow equations, is constructed by the gradient of the action.  Flow equations, however,  are not always obtained by the gradient of the action. For example, the flow equation for the quarks in QCD 
cannot be obtained by the gradient of the action~\cite{Luscher:2011bx}. Also, in the $\mathcal{N}=1$ supersymmetric pure Yang-Mills theory, the flow equation can no longer be a simple gradient of the action and the generalized gradient flow equation has to be introduced~\cite{Kikuchi:2014rla}, in order to keep the supersymmetry.

What do we need for the flow equation ?  Actually, it depends on the purpose. Our ultimate goal is to use the flow equation in order to 
construct the lattice supersymmetry and study its dynamics through Monte-Carlo simulations, following the success for the chiral dynamics in QCD and the energy momentum tensor in Yang-Mills theory. For this goal, we require that the flow equation should respect the supersymmetry 
and should have the remarkable properties of the UV finiteness for the composite operators at finite flow time $t$.  While the former properties can 
be imposed by construction, whether or not the latter property is satisfied can only be seen after studying the resulting equation.  
Since the flow equation may not be unique, the construction of a desirable flow equation may require a trial-and-error process. 
Only after the latter properties are established, one can use it as a good probe for supersymmetric Ward-Takahashi identities. 

In this paper,  
we carry out the construction of the flow equation for a supersymmetric theory as a first step towards the study of lattice supersymmetric  theories, 
taking the $\mathcal{N}=1$ supersymmetric $O(N)$ nonlinear sigma model  in two dimensions as a toy model.  Since this model is analytically solvable 
in the large $N$ limit, one can also address the question  whether the UV finiteness at finite time $t$ is realized or not.
Imposing the condition that the flow equation respect the $O(N)$ symmetry and  SUSY,  and requiring that only four supersymmetric derivative operators
appear in the equation, we can restrict the flow equations to a specific form. Choosing the simplest example as our choice of the flow equation, we give the solution of the equation at the leading order in the large $N$ expansion. We show that the flow equation of the superfield in this model is dominated by the flow of its scalar term, since the SUSY is dynamically unbroken in the large $N$ limit. We show that the two point function in terms of the superfield is finite in the large $N$ limit. 
Although the UV finiteness for operators including sub-leading terms are not shown yet, our finding can give a promising test ground towards the study of lattice supersymmetric theories. 

This paper is organized as follows. In Sec.\ref{sec2} we give general requirements to construct the flow equation. In Sec.\ref{sec3}, using the requirements, we construct the flow equation of  the $\mathcal{N}=1$ supersymmetric $O(N)$ nonlinear sigma model (SNLSM) in two dimensions as a concrete example, which is compatible with the SUSY.  We show the finiteness of the two point function in the large $N$ limit  in Sec.\ref{sec4}. We summarize our results of this paper in Sec.\ref{sec5}. Some general properties for  SUSY in two dimensions  are presented in Appendix \ref{appa}, while the dynamics of SNLSM in two dimensions is discussed in Appendix \ref{app2}.

\section{The Flow Equation} \label{sec2}

We here propose two conditions for the flow equation to satisfy.
\begin{description}
\item[I] Properties of the system are preserved by the flow equation.  \\
In particular, we consider two properties as
\begin{description}
\item[a)] the constraint of the system, 
\item[b)] the symmetry of the system. 
\end{description}
We refer the conditions corresponding to these properties as "I-a" and "I-b" respectively. 
\item[I\hspace{-.1em}I]. The linear part of the flow equation 
for a field $\Omega$
should be given by a diffusion equation as
\begin{eqnarray}
\frac{\partial \Omega}{\partial t}=\Box \Omega+\cdots 
\label{eq:1st_cond}
\end{eqnarray}
to keep the smearing property of the flow equation. This means that the mass dimension of the flow time is $-2$. 
\footnote{
There is also an interesting extention to a wider class of the flow equation with higher derivatives  in $\lambda \phi^4$ theory in Ref.\cite{Fujikawa:2016qis}, 
which we will not discuss in this paper.
}
\end{description}


We here consider the condition II.
If the kinetic part of the action is given by
\begin{eqnarray}
S_0(\Omega)=\int d^D x\, \Omega(x) K(x) \Omega(x),\label{252}
\end{eqnarray}
where $K$ is 
an inverse propagator whose mass dimension is $[K]=D-2[\Omega]$, and the gradient of this action has the mass dimension $D-[\Omega]$, while the mass dimension of the left-hand side (L.H.S.) of eq.~(\ref{eq:1st_cond}) is $[\Omega]+2$.
Therefore, if
\begin{eqnarray}
\left[\Omega\right]=\frac{D-2}{2}\label{condition1}
\end{eqnarray}
is satisfied, the gradient flow equation is given by the gradient of the action. 

The condition \eref{condition1} is satisfied for the Yang-Mills field and the scalar field, while can not be satisfied for the fermion such as  the quark field in QCD~\cite{Luscher:2013cpa}.
Even if the condition \eref{condition1} is satisfied, the condition I sometimes requires the generalized gradient flow equation in order to keep the nonlinearly realized symmetry~\cite{Kikuchi:2014rla}.
The  $O(N)$ nonlinear sigma model (NLSM) is such an example.

If the system has a SUSY, the condition has to be modified slightly. The mass dimension of the 
right-hand side (R.H.S.) of the gradient flow equation given by the variation of the action are shifted by $\mathcal{N}/2$, where $\mathcal{N}$ is a number of supercharges. In fact, the supersymmetric action is provided by
\begin{eqnarray}
S(\Omega)=\int d^D x d^\mathcal{N}\theta \Omega(x, \theta) K(x, \theta) \Omega(x, \theta),\label{252}
\end{eqnarray}
where $[d\theta]=1/2$ and the mass dimension of 
an inverse
super propagator $K(x, \theta)$ is given by 
\begin{eqnarray}
[K]=D-2[\Omega]-\frac{\mathcal{N}}{2} \,.
\end{eqnarray}
Imposing the condition that the mass dimension of the L.H.S. and R.H.S.  of the gradient flow equation match, i.e. 
$[\Omega]+2=[K]+[\Omega]$,  one obtains
\begin{eqnarray}
\left[\Omega\right]=\frac{D-2}{2}-\frac{\mathcal{N}}{4}.\label{condition2}
\end{eqnarray}
Eq.~(\ref{condition2})  is the condition that the flow equation can be obtained by the gradient of the action in the supersymmetric theory,
and it reduces to \eref{condition1} when there is no SUSY($\mathcal{N}=0$).

For a field theory with $\mathcal{N}$ supercharges (including the case $\mathcal{N}=0$), if
the condition \eref{condition2} is not satisfied, the flow equation can not be obtained by the gradient of the action.
While  the flow equation of the NLSM  in two  dimensions without SUSY satisfies the  condition \eref{condition2} with $\mathcal{N}=0$ (i.e. \eref{condition1}), and thus has been extensively studied in Refs.\cite{Kikuchi:2014rla, Makino:2014sta, Aoki:2014dxa, Makino:2014cxa},
its $\mathcal{N}=1$ supersymmetirc extension 
does not satisfy the condition \eref {condition2}, 
so we cannot obtain the flow equation by the gradient of the action. In the next section, we construct the flow equation of the SNLSM in two dimensions which satisfies two requirements I and II.

\section{The Model}\label{sec3}
As a concrete example, we analyze the flow equation of the $\mathcal{N}=1$ 
SNLSM
in two dimensions, whose action is provided by
\begin{eqnarray}
S(\Phi)=\frac{1}{2g^2}\int_{x, \theta}\bar{D}\Phi D\Phi,
\end{eqnarray} 
where the superfield $\Phi(x, \theta)=\varphi(x)+\bar{\theta}\psi(x)+\frac{1}{2}\bar{\theta}\theta F(x)$ is a $N$ components vector field, which satisfies 
\begin{eqnarray}
\sum_{\alpha=1}^{N}(\Phi^{\alpha}(x,\theta))^2=1.\label{constraintt}
\end{eqnarray}
The detailed analysis of the model is given in the Appendix \ref{app2}.



We demand  the condition I-a that the flow equation keeps the constraint.  We extend the constraint \eref{constraintt} to the one in terms of the flowed field at any flow time such that
\begin{eqnarray}
\sum_{\alpha=1}^{N}(\Phi^{\alpha}(t, x,\theta))^2=1.\label{constraintt2}
\end{eqnarray}
If we differentiate both sides of \eref{constraintt2} with respect to the flow time, we obtain
\begin{eqnarray}
\sum_{\alpha=1}^{N}\Phi^{\alpha}(t, x,\theta) \frac{d \Phi^{\alpha}(t, x,\theta)}{dt}=0.
\label{eq:derivative}
\end{eqnarray}
Since this equation means that the LHS of the flow equation of the model is orthogonal to the superfield $\Phi$, the RHS of the flow equation should be proportional to the projection operator such that
\begin{eqnarray}
\frac{d \Phi^{\alpha}(t, x,\theta)}{dt}=\left(\delta^{\alpha\beta}-\Phi^\alpha\Phi^\beta \right)F^{\beta},
\label{eq:flow_SUSY}
\end{eqnarray}
which, together with \eref{constraintt}, leads to \eref{constraintt2} as follows.
We modify \eref{eq:flow_SUSY} a little as
\begin{eqnarray}
\frac{d \Phi^{\alpha}(t, x,\theta)}{dt}=\left(\Phi^2 \delta^{\alpha\beta}-\Phi^\alpha\Phi^\beta \right)F^{\beta},
\label{eq:flow_SUSY_M}
\end{eqnarray}
so that
\begin{eqnarray}
\frac{d}{d t} \sum_{\alpha=1}^{N}(\Phi^{\alpha}(t, x,\theta))^2 &=& 0,
\end{eqnarray}
which implies
\begin{eqnarray}
\sum_{\alpha=1}^{N}(\Phi^{\alpha}(t, x,\theta))^2 &=& \sum_{\alpha=1}^{N}(\Phi^{\alpha}(0, x,\theta))^2 = \sum_{\alpha=1}^{N}(\Phi^{\alpha}(x,\theta))^2=1.
\end{eqnarray}
Since \eref{constraintt2} now holds, \eref{eq:flow_SUSY_M} reduces to \eref{eq:flow_SUSY}.

We also impose the condition I-b that the flow equation retains the supersymmetry, which implies that $F^\alpha$ should be constructed by the super field $\Phi$ as well as the covariant derivative operators $\bar{D}$ and $D$. 
Since these covariant derivative commute with super transformation operator $\bar{\xi} Q$,
then the R.H.S. of  \eqref{eq:flow_SUSY} with $F^\beta(\Phi,\bar{D},D)$ 
 transforms as 
\begin{eqnarray}
F^\beta(\Phi,\bar{D},D) \rightarrow F^\beta(\Phi,\bar{D},D) + \bar{\xi} Q F^\beta(\Phi,\bar{D},D)
\label{eq:F_transf}
\end{eqnarray}
under the infinitesimal super transformation
\begin{eqnarray}
\Phi \rightarrow \Phi + \bar{\xi} Q \Phi.
\end{eqnarray}
This is because 
1) a product of arbitrary superfields $\Phi$ and $\Xi$ is another superfield and $\bar{\xi}Q$ satifsfies the Leibnitz rule
\begin{eqnarray}
\bar{\xi}Q(\Phi \Xi) = \bar{\xi}Q(\Phi) \Xi + \Phi \bar{\xi}Q (\Xi), 
\end{eqnarray}
2) for any superfield $\Phi$, $D_\alpha \Phi$  is another superfield, 
and 3) $Q$ and $D$ anticommutes so that 
\begin{eqnarray}
D_\alpha (\bar{\xi} Q \Phi)= \bar{\xi}Q (D_\alpha \Phi)
\end{eqnarray}
holds. By repeatedly using 1),2),3) one can show that any function $F$ made of $\Phi$, $D_\alpha \Phi$, and higher $D$ derivatives 
obey the transformation rule in Eq.(\ref{eq:F_transf}).
The same property 1),2),3) can also holds for the product of the projection operator
$\left(\delta^{\alpha\beta}-\Phi^\alpha\Phi^\beta \right)$  and the Field $F^\beta$
so that the R.H.S. transforms as
\begin{eqnarray}
\left(\delta^{\alpha\beta}-\Phi^\alpha\Phi^\beta \right)F^{\beta}
\rightarrow \left(\delta^{\alpha\beta}-\Phi^\alpha\Phi^\beta \right)F^{\beta}
+ \bar{\xi}Q \left[\left(\delta^{\alpha\beta}-\Phi^\alpha\Phi^\beta \right)F^{\beta}\right].
\end{eqnarray}
The L.H.S. transforms as
\begin{eqnarray}
\frac{d \Phi^{\alpha}(t, x,\theta)}{dt} \rightarrow  \frac{d \Phi^{\alpha}(t, x,\theta)}{dt} +
\bar{\xi} Q \left(\frac{d \Phi^{\alpha}(t, x,\theta)}{dt}\right),
\end{eqnarray}
if $\xi$ and $\bar\xi$ are $t$ independent. Thus  the flow equation \eqref{eq:flow_SUSY} keeps the supersymmetry.

Since the mass dimension of the super field $\Phi$ must be zero due to \eref{constraintt2}, the mass dimension of $F^\beta$ should be equal to  two.
Let us finally impose the condition I\hspace{-.1em}I \eref{eq:1st_cond}, i.e. the linear part of the flow equation should include diffusion part.
The simplest choice of $F^\beta$, \footnote{The term $(\Phi^\alpha\bar{D}D\Phi^\alpha) \bar{D}D\Phi^\beta$ is also allowed, but we take the most simple choice here.} is given by
\begin{eqnarray}
F^{\beta}=\bar{D}D\bar{D}D\Phi^\beta\,,
\end{eqnarray}
which leads to the flow equation of SNLSM in two dimensions as
\begin{eqnarray}
\frac{d\Phi^\alpha}{dt}&=&(\delta^{\alpha\beta}-\Phi^\alpha\Phi^\beta)\bar{D}D\bar{D}D\Phi^\beta,\label{origin}
\end{eqnarray}
where the superfield $\Phi^\alpha$ 
 automatically satisfies the constraint \eref{constraintt2},
as shown before.
If we solve this constraint, the flow equation becomes 
\begin{eqnarray}
\frac{d\Phi^a}{dt}=(\delta^{ab}-\Phi^a\Phi^b)\bar{D}D\bar{D}D\Phi^b-\Phi^a\sqrt{1-\Phi^2}\bar{D}D\bar{D}D\sqrt{1-\Phi^2},\label{main}
\end{eqnarray}
where we take $\alpha=1,2,\dots,N$ while $a=1,2,\dots,N-1$. After a little algebra and the redeinition of $4t$ to $t$,
we obtain
\begin{eqnarray}
\frac{d\Phi^a}{dt}=\partial^2\Phi^a+\Phi^a \partial\Phi^b\partial\Phi^b+\frac{\Phi^a(\Phi^b\partial\Phi^b)^2}{1-\Phi^2},\label{gf}
\end{eqnarray}
which is identical to the gradient flow equation of the two dimensional $O(N)$ NLSM  
if the superfield $\Phi$ is replaced by the scalar field $\phi$\cite{Kikuchi:2014rla}.

We can also show that the equation is invariant under the global $O(N)$ rotation by transforming both sides of \eref{gf} by operating $\delta$, which is defined by
\begin{eqnarray}
\delta\Phi^\alpha(t,x,\theta) &=&\sum_{\beta=1}^N \omega^{\alpha\beta}\Phi^{\beta}\\
&=&\sum_{b=1}^{N-1} \omega^{\alpha b}\Phi^{b}\pm\omega^{\alpha N}\sqrt{1-\sum_{b=1}^{N-1}\left(\Phi^b\right)^2},
\end{eqnarray}
where $\omega$'s are the infinitesimal parameters for the $O(N)$ rotation.


\section{Results}\label{sec4}
\subsection{Solution to the Flow Equation}
We solve the flow equation 
at the large $N$ limit and examine whether 
the solutions are UV finite or not, using the same method in Ref.\cite{Aoki:2014dxa} . 
The flow equation (\ref{gf}) in the momentum space is written by
\begin{eqnarray}
\frac{d\Phi^a(t,p,\theta)}{dt} &=&-p^2\Phi^a(t,p,\theta)-\int^3_p\Phi^a(t,p_1,\theta)(p_2\cdot p_3)\Phi^b(t,p_2,\theta)\Phi^b(t,p_3,\theta)\nonumber\\
&&-\sum_{n=0}^{\infty}\int_p^{2n+5}\Phi^a(t,p_1,\theta)\frac{p_2+p_3}{2}\cdot\frac{p_4+p_5}{2}\prod_{j=1}^{n+2}\Phi^b(t,p_{2j},\theta)\Phi^b(t,p_{2j+1},\theta),~~~~~~\label{eq}
\end{eqnarray}
where we use the abbreviation as
\begin{eqnarray}
\int_{p}^{n}\equiv\prod_{i=1}^{n}\int\frac{d^2 p_i}{(2\pi)^2}\hat{\delta}\left( \sum_{i=1}^{n}p_i-p\right),~~\hat{\delta}(p)\equiv (2\pi)^2\delta^{(2)}(p).
\end{eqnarray}
Let us consider the flow time dependence of the two point function. In the contraction with any operators, 
the leading term in the large N expansion in \eref{eq} is obtained from self-contraction, so the flow equation is reduced to
\begin{eqnarray}
\frac{d\Phi^a(t,p,\theta)}{dt} &=&-p^2\Phi^a(t,p,\theta)-\int^3_p\Phi^a(t,p_1,\theta)(p_2\cdot p_3)\langle\Phi^b(t,p_2,\theta)\Phi^b(t,p_3,\theta)\rangle\nonumber\\
&&-\sum_{n=0}^{\infty}\int_p^{2n+5}\Phi^a(t,p_1,\theta)\frac{p_2+p_3}{2}\cdot\frac{p_4+p_5}{2}\prod_{j=1}^{n+2}\langle\Phi^b(t,p_{2j},\theta)\Phi^b(t,p_{2j+1},\theta)\rangle,~~~~~~\label{rep}
\end{eqnarray}
which means that the equation is valid in the contraction with any operators at the large $N$ limit. 
Note that the third term of the R.H.S in \eref{rep} vanishes at the large N limit because of the momentum preservation. 

In order to solve \eref{rep}, we employ the following ansatz, 
\begin{eqnarray}
\Phi^a(t,p,\theta)=F(t,\theta)e^{-tp^2}\Phi^a(0,p,\theta),\label{ansatz}
\end{eqnarray}
where $F$ is a superfield function given by $F(t, \theta)=f(t)+\bar{\theta}g(t)+\frac{1}{2}\bar{\theta}\theta H(t)$. 
As shown in  appendix~\ref{app2}, the two point function of the superfield $\Phi$ at $t=0$ and $\theta=\theta'$ 
is given by
\begin{eqnarray}
\langle\Phi^a(0,p,\theta)\Phi^b(0,q,\theta)\rangle=\frac{\kappa}{N}\delta^{ab}\hat{\delta}(p+q)\frac{1}{p^2+m^2}+O\left(\frac{1}{N^2}\right),
\end{eqnarray}
where $\kappa = g^2 N$ is the t'Hooft coupling.
Using this,
we obtain the equation for the superfield $F$ as 
\begin{eqnarray}
\frac{d F(t,\theta)}{d t}=\kappa F^3(t,\theta)I(t,m),~~~F(0)=1,
\label{eq:F_eq}
\end{eqnarray}
where
\begin{eqnarray}
I(t,m)\equiv\int_q q^2 e^{-2q^2 t}\left(\frac{1}{q^2+m^2}\right).
\end{eqnarray}
In the component field expression, \eref{eq:F_eq} reads, 
\begin{eqnarray}
\frac{df(t)}{dt}&=&\kappa f^3 (t) I(t, m),\\
\frac{dg(t)}{dt}&=&0,\\
\frac{dH(t)}{dt}&=&3\kappa f^2(t)H(t) I (t, m),
\end{eqnarray}
with the initial conditions 
\begin{eqnarray}
f(0)=1, g(0)=0, H(0)=0.
\end{eqnarray}
We can easily solve these differential equations as
\begin{eqnarray}
f(t)&=&e^{-m^2 t}\sqrt{\frac{\ln\frac{\Lambda^2+m^2}{m^2}}{\mathrm{Ei}\{-2t(\Lambda^2+m^2)\}-\mathrm{Ei}(-2tm^2)}},\\
g(t)&=&0,\qquad H(t)=0,
\end{eqnarray}
where we use \eref{approximation} and ${\mathrm{Ei}(x)}$ is the exponential integral function defined by
\begin{eqnarray}
{\mathrm{Ei}}(x) =\int_{-\infty}^{x} dt\frac{e^{t}}{t}=-\int^{\infty}_{-x} dt\frac{e^{-t}}{t}.
\end{eqnarray}
Thus the solution of the flow equation finally becomes   
 \begin{eqnarray}
 \Phi^a(t,p,\theta)=f(t)e^{-tp^2}\Phi^a(0,p,\theta).
\end{eqnarray}

A remarkable feature is that the flow time dependence of the fields are common for all components of the superfield. 
This manifestly shows that the flow equations keeps the supersymmetry in the sense that the flow time evolution and supersymmetry transformation commute 
with each other. It is also interesting to see that the scalar component of the solution has the same form as in non-supersymmetric $O(N)$ NLSM.

 \subsection{Finiteness of Two Point Function}
Using the same discussion in Ref.\cite{Aoki:2014dxa}, we show the finiteness of the two point function in terms of the flowed superfield at the leading order in the large $N$ expansion as 
\begin{eqnarray}
\langle\Phi^a (t, p, \theta)\Phi^b (t', p', \theta')\rangle&=&f(t)f(t')e^{-t p^2}e^{-t' p'^2}\langle\Phi^a(0,p, \theta)\Phi^b(0,p', \theta')\rangle\\
&=&f(t)f(t')\frac{\kappa}{N}\delta^{ab}e^{-(t+t')p^2}\hat{\delta}(p+p')\nonumber\\
&&\times \left(\frac{1+\frac{1}{2}m(\bar{\theta}\theta+\bar{\theta}'\theta')-\frac{1}{4}p^2\bar{\theta}\theta\bar{\theta}'\theta'}{p^2+m^2}-\frac{\bar{\theta}(-i\Slash{p}+m)\theta'}{p^2+m^2}\right),~~~~~~~~
\end{eqnarray}
where
the coefficient is given by
\begin{eqnarray}
\lim_{\Lambda\rightarrow\infty} \kappa f(t) f(t') &=& 4\pi \frac{e^{-m^2(t+t')}}{ \sqrt{- {\rm Ei} (-2t m^2)}\sqrt{- {\rm Ei} (-2t' m^2)}},
\end{eqnarray}
which is finite as long as $ t t^\prime \not=0$.
We finally obtain the two point function in terms of the flowed superfield  as
\begin{eqnarray}
\langle\Phi^a(t, p, \theta) \Phi^b(t', p', \theta')\rangle&=& 
\frac{4\pi e^{-(p^2+m^2)(t+t')}\delta^{ab}\hat{\delta}(p+p')}{ N\sqrt{- {\rm Ei} (-2t m^2)}\sqrt{- {\rm Ei} (-2t m^2)}}\nonumber\\
&&\times \left(\frac{1+\frac{1}{2}m(\bar{\theta}\theta+\bar{\theta}'\theta')-\frac{1}{4}p^2\bar{\theta}\theta\bar{\theta}'\theta'}{p^2+m^2}-\frac{\bar{\theta}(-i\Slash{p}+m)\theta'}{p^2+m^2}\right) . ~~~~~~~~~
\label{eq:grad-2pt}
\end{eqnarray}

The fact that the flow equation preserves the supersymmetry and the two point functions (or hopefully n-point functions) is finite can open a possibility of future applications 
to lattice supersymmetries. Although the supersymmetry is violated by the lattice regularization like the continuous translation symmetry,
one may be able to construct the supersymmetric current from the composite operators of  flowed fields in the small time limit,  
following the study of Suzuki \cite{Suzuki:2013gza} for the energy momentum tensor.
Another interesting application is 
to use the flowed fields as a good probe for the supersymmetric Ward-Takahashi identity,
which is used as a measure for the fine-tuning of lattice parameters
in order to recover the supersymmetry in the continuum limit\cite{DelDebbio:2013zaa}. 
These studies are in progress.

\section{Summary and Discussion}\label{sec5}

In this paper, we study requirements for
the  flow equation to satisfy, which  are summarized as follows.
\begin{description}
\item[I] Properties of the system such as the constraint and the symmetry are preserved by the flow equation.
\item[II]  The linear part of the flow equation is given by a diffusion equation. 
\end{description}
On the bases of these requirements, we obtain the flow equation of the $\mathcal{N}=1$ 
SNLSM in two dimensions. The flow equation we constructed has the manifest $O(N)$ symmetry and SUSY, while keeping the constraint $\Phi^2=1$ at any flow time.  We give the solution of the equation at the leading order of the large $N$ expansion.

There are two results in the analysis. 
First of all, the flow of the superfield in the model is dominated by the flow of its scalar term,  since the SUSY is not broken dynamically in the original  theory in two dimensions.
Secondly, we show that the two point function of the superfield is finite at the leading order in the large $N$ expansion. In particular, this result means that the two point function of the fermion field is also finite. 
Although we have so far studied the two point functions only, it is worth mentioning that our study is the first case
to show the finiteness of the two point function for flowed fields in the supersymmetric theory non-perturbatively.
In order to complete the proof for the finiteness of flowed fields, one has to consider the analysis including the sub-leading order in large N expansion\cite{Aoki:2016env}. 



It is important to construct the supercurrent in the lattice field theory, using this SUSY flow equation. 
It is also interesting to analyze properties of other models, e.g. NLSMs with the extended SUSY or ones in the difference dimensions. More general method to construct the flow equation is needed. 
Finally, one may  consider the induced metric discussed in Ref.\cite{Aoki:2015dla, Aoki:2016ohw,Aoki:2016env} using the supersymmetric flowed field analyzed in this paper.

\acknowledgments
The authors would like to thank H.Suzuki, 
S.Yamaguchi, S.Terashima and T.Nosaka for useful discussions. 
S. A. is supported in part by the Grant-in-Aid of the Japanese Ministry of Education, Sciences and Technology, Sports and Culture (MEXT) for Scientific Research (No. JP16H03978), by a priority issue (Elucidation of the fundamental laws and evolution of the universe) to be tackled by using Post K Computer, and by Joint Institute for Computational Fundamental Science (JICFuS).  T.O. is supported in part by the Grant-in-Aid of the Japanese Ministry of Education, Sciences and Technology, Sports and Culture (MEXT) for Scientific Research (No. 26400248).

\appendix

\section{$\mathcal{N}=1$ SUSY in two dimensions}\label{appa}

This appendix summarizes some notations for $\mathcal{N}=1$ SUSY in two dimensions,  based on Ref.\cite{Moshe:2003xn}. 

Integrations 
over the coordinate $x$, momentum $p$ and the supercoodinate $\theta$
are given by 
\begin{eqnarray}
\int_x=\int d^2 x, \ \
&
\displaystyle{\int_p=\int\frac{d^2 p}{(2\pi)^2},} \ \
&
\int_\theta=\int d^2\theta=\frac{i}{2}\int d\theta_2d\theta_1.
\end{eqnarray}
We consider the two dimensional Euclidean theory, whose metric is $\eta_{\mu\nu}\equiv(+,+)$ for $\mu, \nu=1,2$. 
The scalar superfield is 
\begin{eqnarray}
\Phi(x,\theta)=\varphi(x)+\bar{\theta}\psi(x)+\frac{1}{2}\bar{\theta}\theta F(x),
\end{eqnarray}
where $\psi$ and $\theta$ are two component spinors given by
\begin{eqnarray}
\psi=
\left(
\begin{array}{cc}
\psi_1\\
\psi_2
\end{array}
\right)
,
\theta=
\left(
\begin{array}{cc}
\theta_1\\
\theta_2
\end{array}
\right),
\end{eqnarray}
and the Dirac conjugate is defined by
\begin{eqnarray}
\bar{\theta}&\equiv&  {}^{\mathrm{T}} \! \theta \sigma_2
=
\left(\begin{array}{cc}
i\theta_2,&-i \theta_1
\end{array}\right),
\\
\bar{\psi}&\equiv&  {}^{\mathrm{T}} \! \psi \sigma_2
=\left(\begin{array}{cc}
i\psi_2,&-i \psi_1
\end{array}\right).
\end{eqnarray}
 The gamma matrix in two dimensions is given by 
\begin{eqnarray}
\gamma_\mu\equiv\sigma_\mu,
\end{eqnarray}
where $\sigma_\mu$ is the Pauli matrices defined as 
\begin{eqnarray}
\sigma_1=
\left(
\begin{array}{cc}
0&1\\
1&0
\end{array}
\right),
\sigma_2=
\left(
\begin{array}{cc}
0&-i\\
i&0
\end{array}
\right).
\end{eqnarray}
We thus obtain for arbitrary two component spinors $\psi$ and $\chi$
\begin{eqnarray}
\bar{\psi}\chi&=&\bar{\chi}\psi,\\
\bar{\psi}\gamma^\mu\chi&=&-\bar{\chi}\gamma^\mu\psi,\\
\bar{\psi}\gamma^\mu\gamma^\nu\chi&=&\bar{\chi}\gamma^\nu\gamma^\mu\psi.
\end{eqnarray}

The super covariant derivative is defined by
\begin{eqnarray}
D_{\alpha}\equiv\frac{\partial}{\partial\bar{\theta}_{\alpha}}-(\Slash{\partial}\theta)_{\alpha}, \quad \bar{D}_{\alpha}\equiv\frac{\partial}{\partial{\theta}_{\alpha}}-(\bar{\theta}\Slash{\partial})_{\alpha},
\end{eqnarray}
while the supercharge is  given by
\begin{eqnarray}
Q_{\alpha}=\frac{\partial}{\partial\bar{\theta}_{\alpha}}+(\Slash{\partial}\theta)_{\alpha}, \quad \bar{Q}_{\alpha}=\frac{\partial}{\partial{\theta}_{\alpha}}+(\bar{\theta}\Slash{\partial})_{\alpha}.
\end{eqnarray}
We use here the Feynman slash notation,
\begin{eqnarray}
\Slash{\partial}&\equiv&\gamma_\mu\partial_\mu.
\end{eqnarray}
%
%
%
%
Note that 
\begin{eqnarray}
\bar{D}={}^T(\sigma_2 D)
\end{eqnarray}
holds. 
Considering the super covariant derivative of the scalar superfield, we obtain
\begin{eqnarray}
D_{\alpha}\Phi&=&\psi_{\alpha}+\theta_{\alpha}F-(\Slash{\partial}\theta)_{\alpha}(\varphi+\bar{\theta}\psi),\\
\bar{D}_{\alpha}\Phi&=&-\bar{\psi}_{\alpha}-\bar{\theta}_{\alpha}F-(\bar{\theta}\Slash{\partial})_{\alpha}(\varphi+\bar{\psi}\theta),
\end{eqnarray}
Then we obtain
\begin{eqnarray}
\{D_\alpha, \bar{D}_\beta \}&=&-2\Slash{\partial}_{\alpha\beta},\\
\{D_\alpha, Q_\beta \}&=&0.
\end{eqnarray}
It is also easy to see that the following relation holds.
\begin{eqnarray}
(\bar{D}_\alpha D_\alpha )^2=4\partial^2.
\end{eqnarray}

\section{SNLSM in two dimensions in the large $N$ limit}\label{app2}
We consider the $\mathcal{N}=1$ 
SNLSM in two dimensions\cite{Witten:1977xn}. This appendix is the review of Refs.\cite{Moshe:2003xn, Moshe:2002ra}. We derive the same result by the different method. 
\subsection{Action}
The action is 
\begin{eqnarray}
S(\Phi)=\frac{1}{2g^2}\int_{x, \theta}\bar{D}\Phi D\Phi,
\end{eqnarray} 
where the superfield $\Phi(x, \theta)=\varphi(x)+\bar{\theta}\psi(x)+\frac{1}{2}\bar{\theta}\theta F(x)$ is a $N$ components vector field, which satisfies 
\begin{eqnarray}
\sum_{\alpha=1}^{N}(\Phi^{\alpha}(x))^2=1.
\end{eqnarray}
The generating function $Z(J)$ with a source $J$ is given by
\begin{eqnarray}
Z(J)=\int DL D\Phi e^{-S_{tot}(\Phi, L, J)},
\end{eqnarray}
where
\begin{eqnarray}
S_{tot}(\Phi, L, J)&=& \frac{1}{2 g^2}\int_{x, \theta}\bar{D}\Phi D\Phi+\frac{1}{g^2} \int_{x, \theta} L(\Phi^2-1)-\frac{1}{g^2}J\cdot\Phi.  
\end{eqnarray}
Here we introduce
$L(x, \theta)=M(x)+\bar\theta l(x)+\frac{1}{2}\bar{\theta}\theta \lambda(x)$ as the Lagrange multiplier superfield. Integrating out $\Phi$, we obtain
\begin{eqnarray}
Z(J)&=&\int DL e^{-S_{\mathrm{eff}}(L, J)},
\end{eqnarray}
where
\begin{eqnarray}
S_{\mathrm{eff}}(L, J)&=&\frac{N}{2}\mathrm{Str}\log K -\frac{N}{\kappa}\int_{x, \theta} L(x, \theta)-\frac{N}{2\kappa}J \cdot K^{-1}\cdot J,\label{eq: Seff2}~~~\\
K&=&-\bar{D}D+2L,
\label{eq: Seff}
\end{eqnarray}
and $\kappa\equiv g^2 N$ is the t'Hooft coupling.



\subsection{Saddle Point Equation}

Since the effective action $S_{\mathrm{eff}}$ is proportional to $N$, the path integral over the superfield $L$  is dominated by the saddle point of the effective action in the large $N$ limit  as
\begin{eqnarray}
Z(J)&=&e^{-\bar{S}_{\mathrm{eff}}(J)},
\end{eqnarray}
where $\bar{S}_{\mathrm{eff}}(J)={S}_{\mathrm{eff}}(\bar{L}(J), J)$ and $\bar{L}(J)$ is the solution to the saddle point equation given by
\begin{eqnarray}
\frac{1}{N}\frac{\delta S_{\mathrm{eff}}}{\delta L(x,\theta)}&=&\langle x, \theta|(-\bar{D}D+2L)^{-1} |x, \theta\rangle-\frac{1}{\kappa}+\frac{1}{\kappa} (K^{-1}\cdot J)^2\nonumber\\
&=&0.\label{130}
\end{eqnarray}

Here, $\bar{L}(J)$ can be expanded in $J$ as
\begin{eqnarray}
\bar{L}(J)(x,\theta) = L_0(x,\theta) + L_2(x, \theta) +\cdots, 
\end{eqnarray}
where $L_0= \left. \bar{L}\right|_{J=0}$ is the $J$ independent part  and $L_2$ is quadratic in $J$, and so on. 
The two point function at large $N$ is given by differentiating the effective action as
\begin{eqnarray}
\langle \Phi(x,\theta) \Phi(x^\prime, \theta^\prime)\rangle = \frac{\kappa^2}{N^2}\left. \frac{\delta^2 S_{\rm eff}(\bar{L}(J),J)}{\delta J(x,\theta) \delta J(x^\prime,\theta^\prime)} \right|_{J=0}.
\end{eqnarray}
In principle, although the differentiation with respect to $J$ can also act on $\bar{L}(J)$, a straightforward  calculation shows that such contributions cancels between 
the differentiations of the first term and the second term in \eref{eq: Seff2} due to the saddle point equation. Therefore, the propagator turns out to be
\begin{eqnarray}
\langle \Phi(x,\theta) \Phi(x^\prime, \theta^\prime)\rangle = \frac{\kappa}{N}\langle x,\theta | (-\bar{D}D+ 2L_0)^{-1} | x^\prime, \theta^\prime\rangle.
\end{eqnarray}
If there are two or more different solutions to the saddle point equation, the true vacuum can be found as the solution which minimizes the 
effective action at $J=0$.  For this reason, in order to obtain the propagator in the true vacuum, it suffices to know only the $J$-independent part of $\bar{L}$, 
namely $L_0$.

From \eref{130}, $L_0$ is given by the saddle point 
equation
\begin{eqnarray}
 \langle x, \theta|(-\bar{D}D+2L_0)^{-1} |x, \theta\rangle-\frac{1}{\kappa} = 0.
\label{130_2}
\end{eqnarray}
Since the vacuum does not break the translational invariance and Lorentz invariance, $L_0$ is independent of $x$ and can be written as
\begin{eqnarray}
L_0 = M + \frac{1}{2}\bar{\theta}\theta \lambda,
\end{eqnarray}
where $M$ is some constant.


In order to solve the saddle point equation \eref{130_2} for $L_0$ , let us compute the quantity $\tilde{\Delta}_{x,\theta,x^\prime,\theta^\prime}$
for general Lorentz invariant constant background field $L= M + \frac{1}{2}\bar{\theta}\theta \lambda$,
\begin{eqnarray}
\Delta_{x,\theta,x^\prime,\theta^\prime} \equiv \langle x,\theta | (-\bar{D}D + 2 L)^{-1} | x^\prime, \theta^\prime \rangle. 
\end{eqnarray}
From the translational invariance it can be parameterized as
\begin{eqnarray}
\Delta_{x,\theta,x^\prime,\theta^\prime} = \int_k e^{ik (x-x^\prime)} \tilde{\Delta}(k,\theta,\theta^\prime).
\end{eqnarray}
$\tilde{\Delta} (k, \theta, \theta')$ satisfies the relation
\begin{eqnarray}
(-\bar{D}D+2L(\theta))\tilde{\Delta} (k, \theta, \theta')=\delta^2(\theta'-\theta).\label{107}
\end{eqnarray}
We take the explicit form of $\tilde{\Delta}_{x,\theta,x^\prime,\theta^\prime}$ as
\begin{eqnarray}
\tilde{\Delta} (k, \theta, \theta')=a_1+a_2\bar{\theta}\theta+a_3\bar{\theta}'\theta'+a_4\bar{\theta}\theta'+a_5 i \bar{\theta} \Slash{k}\theta'+a_6\bar{\theta}\theta\bar{\theta}'\theta'.
\end{eqnarray}
Using \eref{107}, we obtain coefficients as 
\begin{eqnarray}
a_1&=&\frac{1}{k^2+m^2}, ~a_2=a_3=\frac{\frac{1}{2}M}{k^2+m^2}, ~a_4=-\frac{M}{k^2+M^2}, \nonumber\\
a_5&=&\frac{1}{k^2+M^2}, ~a_6=-\frac{1}{4}\frac{k^2+\lambda}{k^2+m^2},
\end{eqnarray}
where we introduce the boson mass as
\begin{eqnarray}
m\equiv\sqrt{\lambda+M^2}.
\end{eqnarray}
Thus $\tilde{\Delta}_{x,\theta,x^\prime,\theta^\prime}$ is given by
\begin{eqnarray}
\tilde{\Delta} (k, \theta, \theta')=\frac{1+\frac{1}{2}M(\bar{\theta}\theta+\bar{\theta}'\theta')-\frac{1}{4}(k^2+\lambda)\bar{\theta}\theta\bar{\theta}'\theta'}{k^2+m^2}-\frac{\bar{\theta}(-i\Slash{k}+M)\theta'}{k^2+M^2}.\label{superpropagator}
\end{eqnarray}

Now, let us go back to the saddle point equation for $L_0$. 
In momentum representation, the saddle point equation reads
\begin{eqnarray}
\int_k\tilde{\Delta}(k, \theta, \theta)=\frac{1}{\kappa}
\label{161}.
\end{eqnarray}
This equation reduces to 
\begin{eqnarray}
\frac{1}{4\pi} \log\frac{\Lambda^2+m^2}{m^2} &=& \frac{1}{\kappa},\label{approximation}\\
\frac{M}{4\pi} \log\frac{M^2(\Lambda^2+m^2)}{m^2(\Lambda^2+M^2)} &=& 0,
\end{eqnarray}
where $\Lambda$ is the momentum cut-off, and the second equation implies that 1) $M=m$ or 2) $M=0$. 
Using the first equation, the solutions in the large $\Lambda$ limit are given as 
\begin{eqnarray}
M =m= \Lambda e^{-\frac{2\pi}{\kappa}}, \\
M=0, m= \Lambda e^{-\frac{2\pi}{\kappa}}.
\end{eqnarray}

\subsection{Action Density}\label{AD}
We calculate the action density, which is defined by $\epsilon/N=\bar{S}_{\mathrm{eff}}/(NV)$, as 
\begin{eqnarray}
\epsilon/N&=&\frac{1}{8\pi}\left(m^2-M^2 +2M^2\log{\frac{M}{m}}\right).
\end{eqnarray}
Using this result, we find that the true vacuum is realized at 
\begin{eqnarray}
\epsilon=0,~~~M=m.\label{mass}
\end{eqnarray}

\subsection{Super Propagator}
Using \eref{superpropagator} and \eref{mass}, and setting $\theta'=\theta$, we finally obtain the super propagator at large $N$ as
\begin{eqnarray}
\tilde{\Delta}(k, \theta, \theta)=\frac{1}{k^2+m^2}.
\end{eqnarray}

\providecommand{\href}[2]{#2}\begingroup\raggedright\endgroup

\end{document}